\documentclass[aps,pre,twocolumn,superscriptaddress,showpacs]{revtex4}
\usepackage{graphicx}
\usepackage{dcolumn}
\usepackage{bm}
\usepackage{amsmath}
\usepackage{hyperref}
\begin{document}
\title{Nonequilbrium work by charge control in a Josephson junction}

\author{Su Do Yi}
\author{Beom Jun Kim}
\email{beomjun@skku.edu}
\affiliation{Department of Physics and BK21 Physics Research Division,
Sungkyunkwan University, Suwon 440-746, Korea}
\author{Juyeon Yi}
\email{jyi@pusan.ac.kr}
\affiliation{Department of Physics, Pusan National University,
Busan 609-735, Korea}

\date{\today}

\begin{abstract}
We consider a single Josephson junction in the presence of time varying gate charge, and examine the nonequilibrium work done by the charge control in the framework of fluctuation theorems. We obtain the probability distribution functions of the works performed by forward protocol and by its time reversed protocol, which from the Crooks relation gives the estimation of the free energy changes $\Delta F =0$. The reliability of $\Delta F$ estimated from the Jarzynksi equality is crucially dependent on protocol parameters, while Bennett's acceptance ratio method confirms consistently $\Delta F=0$. The error of the Jarzynski estimator either grows or becomes saturated as the duration of the work protocol increases, which depends on the protocol rapidity determining the existence of the oscillatory motion of the phase difference across the junction.
The average of the work also shows similar behaviors and its saturation value is given by the relative weight of the oscillatory trajectory with respect to running trajectories with constant acceleration. We discuss non-negativity of the work average and its relation to heat and entropy production associated with the circuit control.
\end{abstract}

\pacs{ 85.25.Cp, 05.70.Ln, 05.40.-a}

\maketitle

\section{Introduction}

 According to the first law of thermodynamics, the internal energy change occurs due to the work~($w$) done to the system by external forces, and the heat input~($Q$) from the reservoir. Both $w$ and $Q$ are path dependent quantities but the internal energy change remains as a state function. Recently, Jarzynski found that the work alone satisfies a fluctuation theorem~\cite{jarzynski},
\begin{equation}\label{jarzynski}
\int_{-\infty}^{\infty}dw p(w) e^{-\beta w}=e^{-\beta \Delta F}.
\end{equation}
Here the work is performed by applying a time dependent perturbation to a system. Let the control parameter of the perturbation, say $\lambda(t)$, vary in time $t$ according to a prescribed protocol. The work defined by Jarzynski reads
\begin{equation} \label{eq:w}
w=\int_{0}^{T}dt \frac{\partial\lambda(t)}{\partial t}\frac{\partial H}{\partial \lambda},
\end{equation}
where the integration interval corresponds to the time span of the work protocol.
Initially the unperturbed system with $\lambda(0)$ is supposed to be in equilibrium state with a reservoir at the inverse temperature $\beta$,
and hence the initial phase space is populated according to the canonical distribution $e^{-\beta H(\lambda(0))}/Z_{i}$ with the initial partition function $Z_{i}$ related to the initial free energy
$-\beta F_{i}=\ln Z_{i}$.
Due to this randomness of the initial states, the work becomes a stochastic variable that can be characterized by its probability distribution function $p(w)$ which can be acquired by many repetitions of the work protocol. The left hand side of Eq.~(\ref{jarzynski}) is the average of the exponentiated negative work with respect to $p(w)$, which, as the equality indicates, gives the free energy change $\Delta F= F_{f}-F_{i}$ where $F_{f}$ denotes the free energy of the final equilibrium state of the system described by the parameter $\lambda(T)$.
The feasibility of this scheme have been demonstrated in various systems such as single molecules~\cite{sm1,sm2}, classical oscillators~\cite{pen1,pen2,pen3}, and an electronic circuit~\cite{circuit}.

Also the protocol can be performed in bidirectional way. For the forward protocol choosing $\lambda(t)$, we call the work done during this process as the forward work, and write its probability distribution function as $p_{F}(w)$. For the backward protocol, the system is initially in thermal equilibrium with the
perturbation at $\lambda(T)$, and the parameter varies along the time-reversed path of $\lambda(t)$ and $\lambda(t)$ at the end of the backward protocol reaches $\lambda(0)$. We let the probability distribution function for this backward work to be $p_{B}(w)$. These two distributions, $p_{F}(w)$ and $p_{B}(w)$ are related through the Crooks relation~\cite{crooks},
\begin{equation}\label{crooks}
p_{F}(w)=e^{-\beta (\Delta F -w)}p_{B}(-w).
\end{equation}
This relation gives an identity $p_{F}(\Delta F)=p_{B}(-\Delta F)$, indicating that the crossing point between $p_{F}(w)$ and $p_{B}(-w)$ corresponds to $\Delta F$, and hence allows to infer $\Delta F$ without the average process required in Eq.~(\ref{jarzynski}).

In this work, we consider a system of a single Josephson junction and investigate the statistics of the work done by a gate charge control. Although the microscopic origin of involved mechanism in a Josephson junction is of quantum mechanical nature, certain features such as Josephson effects and Shapiro steps are explained purely on the basis of a classical equation of motion for the phase of the condensate wave function~\cite{likarev,tinkham}. We adopt this effective approach and show that the work performed by a time varying gate charge satisfies the fluctuation theorems~(\ref{jarzynski}) and (\ref{eq:w}). This proposes a possibility to verify the fluctuation theorems on a quantum mechanical system yet in the context of classical mechanics. Section~\ref{sec:system} introduces our system and a work protocol to vary the gate charge in time. We consider bidirectional protocols which are related with each other via time reversal operation, and define the Jarzynski work for each direction of the work protocol. We also point out an equivalence to a current controlled junction. In Sec.~\ref{sec:pdf}, we obtain the probability distribution functions~(pdfs) of the forward and the backward work, using Monte Carlo simulation along with the second order Runge-Kutta algorithm. The shapes of pdfs are markedly different depending on the protocol rapidity, and yet a symmetry relation between the forward and the backward work exists to give $\Delta F =0$ according to the Crooks relation.
Sec.~\ref{sec:F} is devoted to the free energy estimation by using the Jarzynski equality, the utility of which is limited by finite size of the work data. The error in the Jarzynski estimation of $\Delta F$ grows rapidly with the time elapse of the protocol, or becomes saturated. In comparison, the Bennett's acceptance ratio method, which is an integrated form of the Crooks relation with a Fermi-function like weighting function introduced~\cite{bennett}, is shown to give reliable estimations of $\Delta F$ irrespectively of the protocol parameters.
Finally in Sec.~\ref{sec:work},  we detail the properties of the work average and its relation to heat and entropy generated by the circuit control, and summary will follow in Sec.~\ref{sec:summary}.

\section{System and work}
\label{sec:system}
We begin with the effective Hamiltonian for a single Josephson junction in the presence of a gate charge $q_{e}(t)$,
\begin{equation} \label{eq:H}
H(t) = \frac{1}{2C}[q(t)-q_e(t)]^2-E_J \cos\theta(t),
\end{equation}
where $\theta$ represents the phase difference of condensed Cooper pairs between the two superconducting grains separated by an insulating barrier. The first term describes a charging energy for excess charge $q-q_{e}$ and the second term denotes the Josephson coupling energy.
We assume that the system is initially in equilibrium with a thermal reservoir at inverse temperature $\beta$ so that the variables at initial time $t=0$, $q(0)$ and $\theta(0)$, are distributed according to the canonical distribution
\begin{equation}\label{can}
P_{eq}(q(0),\theta(0))=e^{-\beta H(0)} /Z_{i}
\end{equation}
with the initial partition function $Z_{i} =
\int d\theta \int dq e^{-\beta H(0)}$.
For the classical picture, the equation of motion for $\theta(t)$ can be obtained from the Hamilton's equation:
$\partial_{t}\theta(t)=(2e/\hbar)\partial_{q}{\cal H}$ and $\partial_{t}q = -(2e/\hbar)\partial_{\theta}{\cal H}$ with
$\partial_{x}=\partial / \partial x$, yielding
\begin{equation}\label{eom}
\partial_{t}^{2} \theta (t)=-2[\sin\theta(t)
+\alpha\partial_{t}q_{e}(t)],
\end{equation}
where the ratio between the Josephson coupling energy and the charging energy
of a single Cooper pair $E_{C}=(2e)^{2}/(2C)$ enters as $\alpha \equiv
\sqrt{E_{C}/E_{J}}$.  In Eq.~(\ref{eom}) and from now on,
the time $t$, the charge, the energy (and the work), and the inverse temperature
are measured in units of $\hbar/\sqrt{E_C E_J}$, $2e$, $E_J$, and $1/E_J$, respectively.
In dimensionless form, the Hamiltonian~(\ref{eq:H}) is written as
\begin{equation}
H(t) = \alpha^2[ q(t) - q_e(t) ]^2 - \cos\theta(t) .
\end{equation}
Equation~(\ref{eom}) equivalently describes the motion of a particle moving in a tilted washboard potential. Depending on the tilting slope, the particle can be either in the locked state staying around the potential minima or in the running state rolling down the hill~\cite{risken}.
In the present problem, the gate charge, $q_{e}(t)$, provides the tilting slope, and its time variance performs the nonequilibrium work on the Josephson junction.

Let us in particular consider the simplest protocols for $q_{e}(t)$. For the forward process, we choose
\begin{equation}\label{proto}
q_{e}(t)=(\gamma/\alpha)t
\end{equation}
 and its time reversed path
$q^{B}_{e}(t)=(\gamma/\alpha)(T-t)$ is taken for the backward process, where the both processes take place for a fixed time span $t = (0,T)$. The parameter $\gamma$ in the protocol controls how rapid the gate charge changes in time.
The Jarzynski work in our system can be obtained by taking
$\lambda(t)=q_{e}(t)$, which for the forward process reads
\begin{equation} \label{eq:workforward}
w=-2\alpha\gamma\int_{0}^{T}dt [q(t)-q_{e}(t)]
=-\gamma [\theta(T)-\theta(0)],
\end{equation}
in units of the Josephson coupling energy $E_J$.
For the second equality, we have used $(1/2\alpha)\partial_t\theta(t) = q(t)-q_{e}(t)$.
For the backward process, denoting the angle path as $\theta^{B}(t)$, we write the work done by the time varying $q^{B}(t)$
\begin{equation} \label{eq:workbackward}
w_{B}=\gamma[\theta^{B}(T)-\theta^{B}(0)],
\end{equation}
in units of $E_J$.
If the backward angle trajectory $\theta^{B}(t)$ is the time reversal of the forward trajectory $\theta(t)$,
that is, $\theta^{B}(t)=\theta(T-t)$, we expect a symmetry relation between the
forward work and the backward work to give $p_{F}(w)=p_{B}(w)$. Combining this
with the Crooks relation Eq.~(\ref{crooks}), we have $p_{F}(w)=e^{-\beta\Delta F}e^{\beta
w}p_{B}(-w)$, which upon inserting $w=0$ gives the free energy change of this
system $\Delta F=0$. This could be expected by that the classical
partition function is invariant under the translation of
the {\it momentum} $q(t)$ by $q_{e}(t)$.
Let us point out that the system described above can be equivalent to a current-biased situation with a special
protocol for varying the bias. In the presence of current bias,
$I(t)$ in units of $(2e/\hbar)\sqrt{E_C E_J}$,
a Josephson junction can be described by
the dimensionless Hamiltonian:
\begin{equation}
H(t)= \alpha^2 q(t)^2-\cos\theta(t) + I(t)\theta(t),
\end{equation}
which again leads to the equation of motion for $\theta$ similar to Eq.~(\ref{eom}):
\begin{equation}\label{eom2}
\partial_{t}^{2} \theta (t)=-2[\sin\theta(t) +I(t)].
\end{equation}
This implies a condition for an equivalence to the charge control: $\alpha
\partial_t q_{e}(t) = I(t)$. For $q_{e}(t)=(\gamma/\alpha)t$, one chooses a
protocol $I(t)=\gamma [\Theta(t)-\Theta(t-T)]$ with $\Theta(x) = 1$ $(0)$
for $x\geq 0$ $(x < 0)$  being
the Heaviside step function, and then, Eq.~(\ref{eom2}) governing the time
evolution of the phase angle $\theta(t)$ becomes identical to Eq.~(\ref{eom})
with the initial distribution of the dynamic variables, $\theta(0)$ and $q(0)$,
also determined by Eq.~(\ref{can}). Furthermore, the work in this case should
be determined by taking $\lambda(t)=I(t)$,
\begin{equation}
w=\int_{0}^Tdt \theta(t)\partial_{t}I(t),
\end{equation}
which is again identical to the work by the gate control (\ref{eq:workforward}). In the foregoing discussion, we only refer to the case of gate charge control and results to be presented can directly be applicable to this current controlled junction.

\section{Probability distribution function of the work and the Crooks relation}
\label{sec:pdf}

\begin{figure}[t]
\includegraphics[width=.45\textwidth]{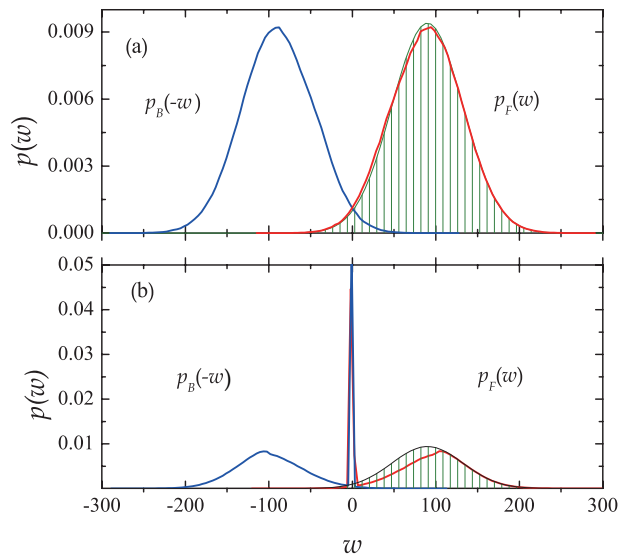}
\caption{Probability distribution functions~(pdfs) of $w$ and $-w_B$~(in units
of $E_{J}$), defined in Eqs.~(\ref{eq:workforward}) and
(\ref{eq:workbackward}), respectively, where we choose the temperature for the
initial thermal equilibrium state of the system, $\beta = 0.1$ (in units of
$1/E_{J})$, and the protocol duration time, $\gamma T=3\sqrt{10}$.  The pdf of
the forward work, $p_{F}(w)$, is mirror symmetric with the pdf of the negative
of the backward work, $p_{B}(-w)$, and their crossing point at $w=0$ gives
$\Delta F=0$ from the Crooks relation.  Panel (a) presents the case for $\gamma=2$,
where the work values are almost normally distributed, as indicated in the
coincidence with the Gaussian distribution function~(the curve with shade) with
average and the variance given by Eq.~(\ref{warunning}).  Panel (b) displays
the work distributions for $\gamma =0.2$ with a sharp peak pronounced around
$w=0$ due to the oscillatory motion of the phase angle. Note that the side peaks are
almost identical to the distribution for $\gamma =2$ apart from the
skewness~(here the shaded region is identical to that presented in the panel
(a)), indicating its origin from the running state agitated at finite
temperatures.
}
\label{fig:crook}
\end{figure}

In order to obtain the pdfs of the work defined by Eqs.~(\ref{eq:workforward}) and (\ref{eq:workbackward}) for the forward and the backward protocols, we first generate the initial dynamic variables, $\theta(0)$ and $q(0)$, according to the canonical distribution Eq.~(\ref{can}). Provided these initial values, the equation of motion (\ref{eom}) is solved through the second order Runge-Kutta method. We acquire $10^{6}$ work values each for the forward and the backward protocol, and present their distributions in Fig.~\ref{fig:crook}. We find that the resulting distribution of the forward work, $p_{F}(w)$, and the distribution of the negative of the backward work, $p_{B}(-w)$ are mirror symmetric about $w=0$, illustrating the symmetry property, $p_{F}(w)=p_{B}(w)$. As anticipated, they cross each other at $w=0$, which from the Crooks relation (\ref{crooks}) indicates $\Delta F =0$. In other words, the perturbation leaves the system free energy unaltered. This result should not depend on the protocol rapidity $\gamma$ in Eq.~(\ref{proto}) which only changes the shape of the distributions. At $\gamma =2$~[see Fig.~\ref{fig:crook}(a)], the pdf of the work appears to be
of the Gaussian form. For large $\gamma$, the nonlinear force term $\sin\theta(t)$ in Eq. (\ref{eom}) can be neglected in comparison with the second term given by $\gamma$ for the protocol (\ref{proto}). For this case, the phase trajectory is approximately given by
$\theta(t)=\theta(0)+v_{0}t+\gamma t^2$ with $v_{0} \equiv \partial_t \theta(t)|_{t=0}$. This running state yields the forward work defined by Eq.~(\ref{eq:workforward}), $w=-\gamma (v_{0}T+\gamma T^{2})$. The average and the second cummulant of this work are obtained as
\begin{eqnarray}\label{warunning}
\langle w \rangle &=& (\gamma T)^{2} \equiv \overline{w}_{r} \\ \nonumber
\sigma^{2}& \equiv &\langle w^{2}\rangle -\langle w\rangle^{2}=2 (\gamma T)^{2}/\beta .
\end{eqnarray}
In Fig.~\ref{fig:crook}(a), we present the Gaussian distribution function given by $P(w)=\exp[-(w-\overline{w}_{r})^{2}/(2\sigma^{2})]/\sqrt{2\pi \sigma^{2}}$~(the curve with shaded area), which coincides well with the numerically obtained $p_{F}(w)$.

On the other hand, when $\gamma$ is small, there exits a locked state around a force-free point, $\theta_{c}=-\sin^{-1}\gamma$. Linearizing Eq.~(\ref{eom}) around $\theta_{c}$, we obtain an oscillatory solution, $\theta(t)-\theta_{c}=A\cos(\Omega t+\phi)$, where the amplitude $A$ and the initial phase $\phi$ are determined by initial conditions, $\theta(0)-\theta_{c}=A\cos\phi$ and
$\partial_t \theta(t)|_{t=0}=2\alpha q(0) =-A\Omega \sin\phi$. Here the oscillation frequency, $\Omega$ is given by
$\Omega^{2}=2|\cos\theta_{c}|$. This oscillatory trajectory yields the work average,
\begin{equation}\label{waosc}
\langle w\rangle = \gamma \theta_{c}[\cos(\Omega T)-1] \equiv \overline{w}_{\ell}~,
\end{equation}
which oscillates in the protocol duration time $T$ with the amplitude much smaller than the average
from the running state [see Eq.~(\ref{warunning})] for $T \gg 1$.
As displayed in Fig.~\ref{fig:crook}(b) for $\gamma=0.2$, this oscillation
results in the sharp peak around $w\approx 0$ in the pdf. Here, the broad side
peak has large overlap with the Gaussian distribution of the running state,
indicating that it originates from the running motion for initial high energy state
at a finite temperature.

\section{free energy estimation from the Jarzynski equality and the Bennett method}
\label{sec:F}

We have shown that the probability distribution functions $p_{F}(w)$ and
$p_{B}(-w)$ obtained from the bidirectional protocols cross with each other at
$w = \Delta F=0$, illustrating the feasibility of the Crooks relation in
determining the free energy change caused by the time varying gate charge in a
single Josephson junction.
This should be confirmed also from the Jarzynski identity.
Within the limitation on the number of work measurements ($N=10^{6}$ in this study),
the estimation of free energy difference $\Delta F$ can be made from
\begin{eqnarray}\label{finiteJ}
\beta \Delta F_{F} &=& -\ln\left[\frac{1}{N}\sum_{i=1}^{N}e^{-\beta w_{i}}\right],
\\ \nonumber
\beta  \Delta F_{B} &=& \ln\left[\frac{1}{N}\sum_{i=1}^{N}e^{-\beta w_{i,B}}\right],
\end{eqnarray}
where $w_{i}$ and $w_{i,B}$ denote the $i$th realization of the work in the forward and the backward protocol, respectively. Since the backward work is symmetrically distributed with respect to the pdf of the forward work, as exemplified in Fig.~\ref{fig:crook}, we have $\Delta F_{B} \approx - \Delta F$, and finite $\Delta F$ indicates the bias of the
estimation due to the finiteness of sample size $N$.

It is often the case that the Jarzynski equality fails to give a reasonable
estimation of $\Delta F$ because the exponential work average crucially
depends on the sampling of the tail part the work distribution with $w
\lesssim \Delta F$~\cite{jarzynski2, fox}. The finite sampling error in the
large sampling limit was investigated in several
studies~\cite{bias1,bias2,zucker1,zucker2}.  Probing the region $w \lesssim
\Delta F$ is also important in extracting $\Delta F$ from the Crooks relation
for the crossing criterion between $p_{F}(w)$ and $p_{B}(-w)$. In our case, the
exact value of free energy change is $\Delta F=0$, and therefore, the
reliability of the Jarzynski identity and the Crooks relation depends on how
low the probability to observe the work $w\lesssim 0$. For the Gaussian
distribution displayed in Fig.~\ref{fig:crook} (see the upper panel), upon
increasing $\gamma T$, the center of the distribution given by the
average~(\ref{warunning}) moves toward large positive values, and  $p(w\lesssim
0)$ becomes extremely small. In this case, the Jarzynski estimation can possess a
severe error, and also the Crooks relation likely fails to give $\Delta F$ due to
the absence of the overlap region between $p_{F}(w)$ and $p_{B}(-w)$.

\begin{figure}
\includegraphics[width=.39\textwidth]{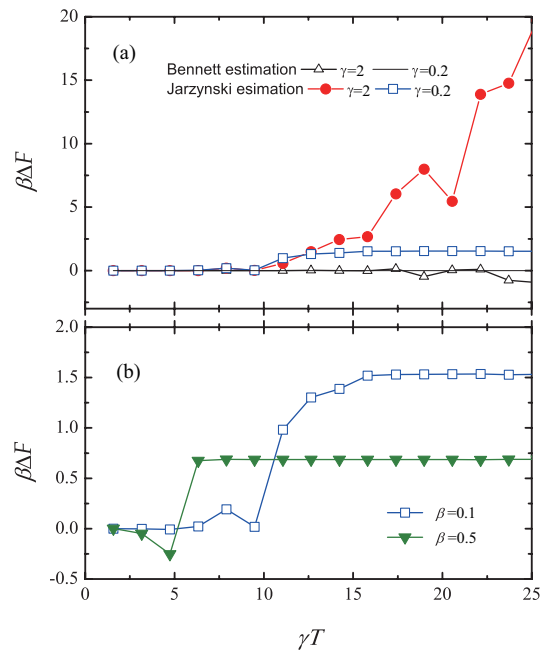}
\caption{(a) The bias of the free energy change calculated by using Eq.~(\ref{finiteJ}) and
the Bennett's acceptance ratio method as a function of the protocol parameter, $\gamma T$, at the inverse temperature
$\beta =  0.1$ (in units of $1/E_J$).
The Bennett method~(line for $\gamma =0.2$ and triangles for $\gamma=2$) persistently yields $\Delta F \approx 0$. but the Jarzynski estimation gives incorrect values of $\Delta F$ for $\gamma T \gtrsim 10$. The error for $\gamma =2$ (filled circles) rapidly grows as $\gamma T$ increases, whereas for $\gamma =0.2$~(open squares) the estimated $\Delta F$ becomes saturated. Panel (b) details this saturation behavior, which also shows the temperature dependence; the saturated error is smaller at lower
temperature~($\beta =0.5$).}
\label{fig:bennett}
\end{figure}

It is well worth introducing another method which is less restrictive than those, called the Bennett's acceptance ratio method based on the following equation~\cite{bennett}:
\begin{equation}\label{bennett}
\int_{-\infty}^{\infty} dw \frac{p_F(w)}{1+e^{\beta(w-\Delta
F)}}=\int_{-\infty}^{\infty} dw \frac{ p_B(-w)}{1+e^{\beta(\Delta F-w)}},
\end{equation}
which can be obtained by multiplying the both sides of Crook's relation (\ref{crooks}) by a Fermi function
$
1/(1+e^{\beta(w-\Delta F)})
$.
This equation was originally implemented in order to estimate partition function ratios by means of Monte Carlo sampling, where the Fermi-function weight in Eq.~(\ref{bennett}) was derived from the requirement of minimal variance of the partition function estimator in the large sample limit. Recently, it was shown by Shirts et al.~\cite{shirts} that a maximum likelihood estimate of the free energy change also yields that Bennett's acceptance ratio method. In various studies, the estimation of $\Delta F$ from Eq.~(\ref{bennett}) was demonstrated to outperform the Jarzynski method and Crooks' crossing criterion~\cite{P,Y,N,Yi}.
For an equal size $N$ of forward and backward samplings, Eq.~(\ref{bennett}) is written as
\begin{equation}\label{finiteB}
\sum_{i=1}^{N} \frac{1}{1+e^{\beta(w_{i}-\Delta F)}} =\sum_{i=1}^{N}\frac{1}{1+e^{\beta(\Delta F+w_{i,B})}},
\end{equation}
which  yields the free energy estimation $\Delta F$ from the Bennett's acceptance ratio method.
In practice, especially when $\gamma$ is large
we observe that the values of the left- and the right-hand sides of Eq.~(\ref{finiteB}) become extremely small
below the numerical accuracy of computer in a broad range of $\Delta F$. In this case, a
precise estimation of $\Delta F$ from Eq.~(\ref{finiteB}) is not plausible, and we instead
approximately estimate $\Delta F$ as $(\Delta F_F -\Delta F_B)/2$~\cite{Yi}, where $\Delta F_F$ and
$\Delta F_B$ are as defined in Eq.~(\ref{finiteJ}).

In Fig.~\ref{fig:bennett}, we present results for the free energy estimations
from the Jarzynski relation and the Bennett method.  As shown in the upper
panel of Fig.~\ref{fig:bennett}, estimations based on the Bennett method are
very close to the true value~($\Delta F =0$) almost insensitively of the
protocol parameter, $\gamma T$. Meanwhile, the Jarzynski estimations have
finite errors. For $\gamma =2$ [see the filled circles in
Fig.~\ref{fig:bennett}(a)], in particular, the error becomes more significant
as $\gamma T$ increases. In fact, in order to obtain a relatively reliable
estimation of $\Delta F$ from the Jarzynski equality, the number of acquired
work values should be larger than $N_{c}\sim e^{\beta h}$ with the hysteresis,
$h=(\langle w\rangle +\langle w_{B}\rangle)/2$~\cite{jarzynski2}.
From this one can estimate the onset point of the error by solving $N=N_{c}$;
\begin{equation}
\beta (\gamma T)_{c}^{2} = \ln N,
\end{equation}
where we used $h\approx \overline{w}_{r} = (\gamma T)^{2}$. For the sample size, $N=10^{6}$, and the
inverse temperature $\beta =0.1$, we find $(\gamma T)_{c}\approx 12$ at which indeed, the Jarzynski error for $\gamma=2$ in Fig.~\ref{fig:bennett}(a) begins to rise rapidly. Meanwhile, the error for $\gamma =0.2$
[see the open squares in Fig.~\ref{fig:bennett}(a)] is much smaller than for $\gamma =2$, and shows a saturation. As remarked, when $\gamma < 1$, there exists an localized motion of the phase variable near the potential minima which yields the work $w\approx 0$, as manifested by the sharp peak at the origin of the pdf in Fig.~\ref{fig:bennett}(b).
These work values make the dominant contribution to Eq.~(\ref{finiteJ}) where the contributions from positively large $w$ populated in the side broad peak for the running motions are exponentially small. These work values pinned near the origin are responsible for the smaller bias than the case of $\gamma > 2$, and also for the saturation. Roughly, we take the work values from the localized oscillatory motions to be $w=0$ and let $N_{\ell}$ to be the number of such realizations of the work values. Then we have
\begin{equation}\label{frac}
\beta \Delta F \approx -\ln (N_{\ell}/N),
\end{equation}
which tells that the saturated value of $\Delta F$ is the fraction of the
localized trajectories relative to the running motions. This fraction depends
on the temperature, as shown in Fig.~\ref{fig:bennett}(b), where at the lower
temperature $\beta =0.5$, the saturated value of $\Delta F$ becomes
smaller, well corresponding to an expectation that the running state must be
more suppressed at low temperature. The relative fraction between the localized
and the running motion comes into play also in the work average, which will be
discussed in the next section.

\section{work average and entropy production}
\label{sec:work}
\begin{figure}[t]
\includegraphics[width=.39\textwidth]{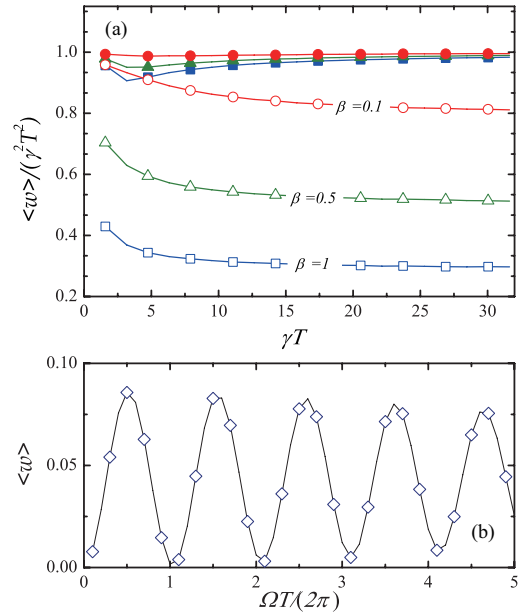}
\caption{(a) The work average is depicted as a function of the total time
elapse $T$ of the work protocol at various temperatures, where the ordinate is
chosen to be the work average divided by $(\gamma T)^{2}=\overline{w}_{r}$ in Eq.~(\ref{warunning})
corresponding to the work
average by the running state. The values for $\gamma =2$ (filled symbols) are
very close to unity, showing the dominance of the running state, at any
temperatures presented here~(temperatures are denoted in the lines of
corresponding empty symbols for $\gamma =0.2$). For $\gamma=0.2$,  the work
average appears to be saturated into a certain fraction of $\overline{w}_{r}$
which becomes reduced by lowering temperature (i.e., increasing $\beta$). Panel
(b) shows the oscillation of the work average for $\gamma =0.2$ at extremely
low temperature $\beta =10$, which reaches the minimum value when the
protocol duration $T$ is close to an integer multiple of the phase oscillation
period, $2\pi / \Omega$.}
\label{fig:work}
\end{figure}

Another important consequence of the fluctuation theorems is the second law of thermodynamics. The Jarzynski equality, $\langle e^{-\beta w}\rangle = e^{-\beta \Delta F}$ together with the Jensen's inequality, $e^{-\beta \langle w\rangle} \leq \langle e^{-\beta w}\rangle$ lead to $\langle w\rangle \geq \Delta F$. Let us remind of relations, $\Delta F = \Delta U - \Delta S/(k_{B}\beta)$ and $\langle w\rangle-\Delta U=\langle Q\rangle$. Here $Q$ is the heat absorbed by the reservoir during the equilibration process; the system at the end of the work protocol is in nonequilibrium state and can reach the final equilibrium sufficiently long after it is brought into a thermal contact with a reservoir.
Then, the inequality, $\langle w\rangle -\Delta F \geq 0$ can be translated into $\Delta S +\Delta S_{r} \geq 0 $ if the reservoir absorbs the heat $Q$ reversibly from the system so that the entropy change of the reservoir is given by $\Delta S_{r}=k_{B}\beta\langle Q\rangle$~\cite{entropy1,entropy2}. Hence, the total entropy production in the system plus reservoir is always positive: $\Delta S_{tot}=\Delta S +\Delta S_{r} \geq 0$. For $\Delta F=\Delta U=0$ as in our case studied here, the work average, identical to the average of the heat, determines the total entropy production:
\begin{equation}\label{ent}
\langle w\rangle = \langle Q\rangle = \Delta S_{tot}/(k_{B}\beta)\geq 0~.
\end{equation}
Note here that the positivity of $\langle w\rangle$ for the second law of thermodynamics also indicates the direction of the heat from the system into the reservoir.

Having this in mind, let us examine the work average. As noted in Sec.~\ref{sec:F}, the work average
crucially depends on $\gamma$. For $\gamma > 1$, $\theta(t)$ increases in time, so does the work average. If nonlinear effect due to the sinusoidal force can be neglected for $\gamma \gg 1$, $\overline{w}_{r}$ for the running state given in Eq.~(\ref{warunning}), can be a good approximation of the work average. In Fig.~\ref{fig:work}(a), we present the work average for $\gamma = 2$ as a function of the time elapse of the work
protocol. In order to see more clearly the convergence behavior of $\langle w\rangle$ into $\overline{w}_{r} = (\gamma T)^{2}$, we choose the ordinate to be $\langle w\rangle/(\gamma T)^2$.
On the other hand, for $\gamma < 1$, in addition to the running state, the phase motion can also be localized around $\theta_{c}=-\sin^{-1}\gamma$ and yields the work average oscillating in time, as discussed in Sec.~\ref{sec:pdf}. In this case, the work average can roughly be
$\langle w\rangle \approx (N_{r}/N)\overline{w}_{r}+ (N_{\ell}/N)\overline{w}_{\ell}$ with $N_{r}=N-N_{\ell}$.
In particular for $\gamma T \gg 1$, since $\overline{w}_{r}\gg \overline{w}_{\ell}$, we have that
 $\langle w\rangle \approx (N_{r}/N)\overline{w}_{r}$. Therefore, the saturated value of
the work averages shown in Fig.~\ref{fig:work}(b) for $\gamma =0.2$ gives
$N_{r}/N$. At $\beta =0.1$, we have $N_{r}/N \approx 0.8$,
which is consistent with $N_{\ell}\approx 0.2$ estimated from the saturation
of $\Delta F\approx 1.5 $ [see Fig.~\ref{fig:bennett}(b)] given in Eq.~(\ref{frac}). This
fraction of the running state decreases as lowering the temperature. At $\beta =0.5$,
we have $N_{r}/N \approx 0.52$, well corresponding to $\Delta F =
0.74$ [see Fig.~\ref{fig:bennett}(b)] giving $N_{\ell}/N \approx 0.47$.
At the extremely low temperature, the contribution from the running state becomes vanishingly small and therefore, the oscillating behavior becomes clearly visible, as displayed in Fig.~\ref{fig:work}(b) for $\beta =10$.
Recalling here the relation Eq.~(\ref{ent}) stating that the work average is identical to the amount of the heat transferred from the system into the reservoir, and the entropy production, one finds that the heat and entropy generation associated with the circuit control crucially depends on $\gamma$ which determines the protocol rapidity by $\partial_t q_{e}(t) = \gamma/\alpha$.
For a fixed value of $\alpha$ and the terminal value of the gate charge $q_{f}\equiv q_{e}(T)=\gamma T/\alpha$, the heat generation for a fast protocol ($\gamma > 1$) is given by
$\langle Q\rangle \approx \overline{w}_{r}= \alpha^2 q_{f}^{2} = (E_C/E_J) q_f^2$ (in units of $E_J$),
corresponding to the energy required to charge the junction of the capacitance $C$ with the terminal gating charge $q_{f}$. For a slow protocol giving $\gamma < 1$,
the heat generation in the large $\gamma T$ limit can be expressed as
$\langle Q\rangle \approx \alpha^2({\widetilde q}_{f})^{2}$ with an effective gate charge
${\widetilde q}_{f}\equiv \sqrt{N_{r}/N}q_{f}$ modified by the temperature dependent factor $N_{r}/N$, which can be
significantly suppressed by lowering temperature.

\section{summary}
\label{sec:summary}

In summary, we considered a single Josephson junction and examine the
nonequilibrium work done by a time varying gate charge. We found that the
details of the probability distribution function of the work depends on the
rapidity of the protocol. For a rapid protocol~($\gamma > 1$), the running
state is dominant to yield the work values which are normally distributed. A
slow protocol~($\gamma < 1$) allows the locked state which manifests itself by
a pdf with sharp peak near the origin, which at finite temperature becomes
bimodal with a broad side peak from the running state. A symmetry relation
between the forward and the backward distribution, $p_{f}(w)=p_{B}(w)$, exist
to yield $\Delta F =0$ according to the Crooks relation. The Jarzynski
estimation of $\Delta F$ can be biased for the finite sample size of the work data.
The error grows rapidly with the time elapse if $\gamma > 1$, and for a slow
protocol $\gamma < 1$ the allowed oscillatory motion saturates the error into a
certain value which
is determined by the relative population of the locked state with respect to the running state.
The Bennett's acceptance ratio method gives reliable estimation, $\Delta F \approx 0$, irrespectively of the protocol parameters. We also discuss the behaviors of the work average reflecting the two characteristic motions, in relation to the heat and the entropy production associated with the charge control.

\acknowledgments
B.J.K. was supported by the National Research Foundation of Korea(NRF) grant funded by the Korea government(MEST) (No. 2010-0008758). J.Y. acknowledges support from the National Research Foundation of Korea (NRF) grant funded by the Korea government (MEST) (No. 2011-0021296).

\end{document}